\documentclass[preprint,aps,tightenlines,floatfix,showpacs]{revtex4}

\usepackage{graphics}
\usepackage{bm}
\usepackage{amsmath}
\usepackage{amssymb}

\begin{document}

\title{Microscopic model analyses of proton scattering from ${}^{12}$C, 
${}^{20}$Ne, ${}^{24}$Mg , ${}^{28}$Si and ${}^{40}$Ca}

\author{Y. J. Kim}
\email{yjkim@cheju.ac.kr} 
\affiliation{Department of Physics,
Cheju National University, Jeju 690-756, Republic of Korea}

\author{K. Amos}
\email{amos@physics.unimelb.edu.au} 
\affiliation{School of
Physics, The University of Melbourne, Victoria 3010, Australia}

\author{S. Karataglidis}
\email{S.Karataglidis@ru.ac.za} 
\affiliation{Department of Physics
and Electronics, Rhodes University, Grahamstown 6140, South Africa}

\author{W. A. Richter}
\email{richter@sun.ac.za}
\affiliation{Department of Physics, University of the Western Cape,
Private Bag X17, Bellville 7530, South Africa,}

\date{\today}

\begin{abstract}

Differential cross sections and analyzing powers for elastic  scattering
from, and for inelastic proton scattering to a set of $2^+_1$ states in,
${}^{12}$C, ${}^{20}$Ne, ${}^{24}$Mg, ${}^{28}$Si and ${}^{40}$Ca,   and
for  a set of energies between 35 to 250 MeV, have been analyzed.      A
$g$-folding model has been used to determine optical potentials  and  a 
microscopic distorted wave approximation taken to analyze the  inelastic 
data.     The effective nucleon-nucleon interactions used to specify the 
optical potentials have also been used as the transition    operators in
the inelastic scattering processes.                Shell and large space 
Hartree-Fock models of structure have been used to describe the  nuclear 
states.
\end{abstract}

\maketitle

\section{Introduction}

Proton scattering is a most useful means to study both  macroscopic
and microscopic aspects of   nuclear   structure  especially now that a 
coordinate  space  model analysis of proton-nucleus scattering has been 
established as a pertinent means to analyze scattering data  from  most 
stable nuclei and for energies in the range $\sim 30$ to $\sim 300$ MeV.
The model as used currently~\cite{Am00} enables predictions to be  made  
of differential cross sections as well as  of spin observables.    When 
good models of structure are used,  those  predictions agree  well with 
data~\cite{Am00}.  Using the relative motion wave functions so found as 
distorted waves in a distorted wave approximation (DWA)  for  inelastic
scattering, predictions have been  made that also agree well       with
observed  results.  Again  that  is  so  when good prescriptions of the 
structure for the inelastic transitions is used.

Elastic scattering, the predominant event caused by  the interactions of 
nucleons with nuclei, has been studied  extensively   over many decades. 
Invariably  data  are  compared  with  results  calculated  from optical 
potentials. All formulations of the nuclear optical model have in common 
an allowance of flux loss from the incident beam to non-elastic channels 
when energies are above reaction thresholds.      By far the most common
approach has been phenomenological with a local, parameterized, 
potential. 
Customarily that is of Woods-Saxon form.  Much effort has gone into
establishing energy and mass dependences of  the  parameter  values with
which good fits to data are found.      Such a process though has hidden 
dangers when a physical interpretation is sought for the potentials, and 
especially when the associated relative motion  wave  functions  through 
the  nuclear  volume  are used  in a DWA to study non-elastic reactions. 
First,  fits   to   elastic   scattering   data require definition of an 
appropriate  set  of  phase  shifts  which only require solutions of the
partial wave Schr\"odinger equations with the optical potentials   to be
defined reasonably asymptotically, i.e. at very large distances from the
nucleus. Such do not test the credibility of the wave functions  through
and near to the nucleus itself.  Second, all phenomenological potentials
inherently support spurious states which, in the actual nucleus,     are 
either Pauli blocked or hindered.  Such can be simply ignored unless the
local interactions form part of a coupled-channel problem.      Then the
violation    of    the    Pauli    principle    has    most      serious
consequences~\cite{Ca05,Am05}. Third, due to the indistinguishability of
the nucleons, knock-out processes exist,   where the detected nucleon is
not the incident one.  This property (also of the Pauli principle)  make
a nucleon-nucleus ($NA$) optical potential very non-local.           The
relative wave functions found from non-local potentials and those 
obtained from  any local form that is phase equivalent are very 
different.   Finally, only
with allowance of such non-locality can the  determined  interaction  be
related to structure properties of the target.  At the least the optical
potentials reflect the full one-body density matrix elements (OBDME)  of 
the nuclear states and not  just  the  matter  densities of the targets.
Similar concerns  exist with  many  calculations of inelastic scattering 
reported in the literature.      Those failing to allow for the exchange 
scattering amplitudes,        which includes almost all  coupled-channel 
calculations built upon collective models, can only achieve the momentum
transfer  variations seen in data by judicious manipulation of parameter
values. Such, by the commission of a  violation of  the Pauli principle, 
have no true physical significance. 

An objective of $NA$ data analyses,  and undoubtably the prime one,   is
to use the calculated results in comparison  with  data  to  assess  the
quality of the nuclear structure model assumed.   Additionally one hopes
to identify mismatches and to understand why they  occur. To achieve 
those hopes, the reaction model used must be as sophisticated  as  the  
structure  one  to be
tested. The Pauli principle must be preserved and its prime consequences
evaluated as fully as possible. Those of the $g$-folding and DWA methods
in current use~\cite{Am00,St02,Ka02} seem appropriate to the task.  With 
them, analyses of data from the scattering of  radioactive  Helium  ions 
from hydrogen~\cite{St02} showed that ${}^6$He has  an  extended neutron
distribution that has been called a neutron halo while ${}^8$He has just 
a skin of neutrons.    Likewise,  on  using detailed structure models of
${}^{208}$Pb,   the  $g$-folding  methods  revealed~\cite{Ka02} that the 
nucleus should have a  neutron  root  mean  square radius $\sim 0.16$ fm 
larger than that for protons,           consistent with predictions from 
evaluations of the neutron equation of state~\cite{Br00x}.

    Microscopic optical potentials built using the $g$-folding model and
with  effective  nucleon-nucleon ($NN$)  interactions  in  the   nuclear
medium, have been used successfully to describe $NA$ elastic  scattering
of nucleons with  energies  in the range $\sim 30$ to $\sim 300$ MeV and
for targets over the  whole range of mass~\cite{Am00}. When good nuclear
structure  details  are  used in the foldings,  no  {\it  a  posteriori}
adjustment  to  those  potentials are required to find a credible  match
to data.  The structure details required are the OBDME, which are to  be 
found  from  large  space  nucleon  models  of the nucleus, and a set of 
single-particle bound-state wave functions. The latter may be those used 
in  the  self-same  structure  model  calculations,  or  selected   from 
consideration of other properties of the nucleus,  such as the root mean 
square radii or electron scattering form factors.

In this paper, by use of the methods described above,   we analyze 
the differential cross sections and analyzing powers for proton  elastic 
scattering from,   and  inelastic  proton  scattering to $2^+_1$ excited 
states in,   ${}^{12}$C,   ${}^{20}$Ne,   ${}^{24}$Mg,  ${}^{28}$Si, and 
${}^{40}$Ca, for  a range of energies from 35 to 250 MeV. We use 
a   shell   model  (SM)~\cite{Ka96}   and   a  Skyrme-Hartree-Fock (SHF) 
model~\cite{Br00,Ri03} to describe the  ground-state  structure of these
nuclei.     The single-particle bound-state wave functions that complete
the nuclear density matrices are   harmonic oscillator (HO), Woods-Saxon
(WS), or as given  by  the     SHF studies~\cite{Br00,Ri03}.     For the
inelastic scattering calculations,   the no-core shell model was used to
define the  transition OBDME for the  excitation of the $2^+_1$ state in
${}^{12}$C.   For $2^+_1$ transitions in the $sd$-shell nuclei, however,
only projected Hartree-Fock (PHF)   studies~\cite{Ne77,Ne79}  have  been
made in a large enough basis to suit.       There is an inconsistency in
using SHF wave functions with PHF transition OBDME,      but we can only
await SHF studies giving    excitation   spectra   and transition OBDME.
However, with such structure, good results for electron scattering  form 
factors were found~\cite{Ka95}.

The DWBA98 computer code of Raynal~\cite{Ra98} has been used to evaluate
both elastic and inelastic (DWA) scattering observables,   in  which  an
effective $NN$ interaction (the Melbourne interaction) can be,   and has
been, used successfully.     The DWBA98 program explicitly evaluates the
knock-out (exchange)  amplitudes  that are a result of the  (two-nucleon
state)  antisymmetry  requirement    to  conserve  the  Pauli principle.

The paper is arranged as follows. In the next section, for completeness,
salient  details  of  microscopic  scattering  theories  to  analyze the 
elastic    and    inelastic    scattering    data  are  given.   Also in 
Sec.~\ref{theory}, the structure information required, and determined by 
diverse models of structure, are discussed.                     Then, in 
Sec.~\ref{Results}, we present and discuss the results of our   analyses 
of the cross sections and analyzing powers.      Conclusions we draw are 
given thereafter in Sec.~\ref{Conclusions}. 


\section{Nucleon-nucleus scattering and nuclear structure}
\label{theory}

  As indicated above,  we  have used a $g$-folding model for the optical
potentials with which predictions of elastic scattering observables have 
been made.    We give relevant details of that model in the first of the 
subsections that follow.    Those potentials then have been used to find 
the distorted waves in DWA calculations of inelastic scattering and some 
details of the scattering amplitudes that are evaluated  are  given   in 
subsection~\ref{DWAform}.      Only salient features are given since the 
theories have been elucidated quite fully in a review~\cite{Am00}. Then, 
in subsection~\ref{structures}, we describe the structure models used to 
provide the information required in those scattering theories,  again in 
brief as they have been fully explained previously~\cite{Br00,Ri03}.

\subsection{$g$-folding model of optical potentials}
\label{g-fold}
  With this model, a microscopic complex, nonlocal, and energy-dependent
optical potential is obtained from folding an effective $NN$ interaction 
with the OBDME,  which  for  the  moment  we  take   to include the wave
functions of the individual bound nucleons, determined from a nucleon-based
model of the structure of the nuclear target state.   While three-
(and more)-body effects are not considered explicitly in this  approach, 
many-nucleon correlations are part of the  structure  calculations  and, 
inherently, in the scattering calculations through the medium dependence 
of the effective $NN$ interactions used.

    In coordinate space the $g$-folding optical potential can be written 
as~\cite{Am00}
\begin{equation}
U({\bf r_1},{\bf r_2};E)  
= \delta({\bf r_1}-{\bf r_2}) \sum_{n}
\zeta_{n} \int \varphi^\ast_{n}({\bf s}) v_{D}(R_{1s})
\varphi_{n}({\bf s})\; d{\bf s}  
+  \sum_{n}\zeta_{n}{}
\varphi^\ast_{n}({\bf r_1}) v_{Ex}(R_{12}) \varphi_{n}({\bf r_2})
, 
\label{app_locton}
\end{equation}
where  $D$  and  $Ex$ denote the sets of elements of the $NN$  effective
interaction that define direct and exchange parts  of the  $NA$  optical
potential.   In this form,  $\zeta_{n}$ are the shell occupancies in the 
target state,   though  more  generally  they are the OBDME if there are 
non-Hartree contributions possible.       $\varphi_{n}({\bf r})$ are the 
single-nucleon bound-state wave functions.

The direct term in Eq.~(\ref{app_locton}) is the well-known   $g\rho$ 
form of the optical potential,
\begin{equation}
V_D({\bf r_1}) = \delta({\bf r_1}-{\bf r_2}) \sum_{n}
\zeta_{n} \int \varphi^\ast_{n}({\bf s}) v_{D}(R_{1s})
\varphi_{n}({\bf s})\; d{\bf s}  
= \delta({\bf r_1}-{\bf r_2}) 
\int \rho({\bf s})\, v_{D}(R_{1s})  d{\bf s} .
\end{equation}
It is local by definition.   Nonlocality of the optical potential arises 
from the explicit exchange terms,  neglect  of which can lead to serious 
problems.  Localization of these non-localities,     even if one forms a 
phase equivalent interaction, is no panacea.      Of course, to follow a
better approach  requires  use  of  credible nucleon-based models of the
target structure.

\subsection{Proton inelastic scattering}
\label{DWAform}
 Inelastic scattering calculations have been made using the DWA with the 
effective $NN$ interaction  taken in the folding to define the   optical 
potential as the transition operator.

With   ${\cal A}_{01}$    being  a  two-nucleon state antisymmetrization 
operator, the transition  amplitudes  for   nucleon inelastic scattering
through   a  scattering angle $\theta$ and between states $J_i, M_i$ and
$J_f, M_f$ in a nuclear target, have the form ~\cite{Am00}
\begin{eqnarray}
{\cal T} &=& T^{M_fM_i\nu^\prime\nu}_{J_fJ_i}(\theta)
\nonumber\\
&=&\left\langle \chi^{(-)}_{\nu^\prime}({\bf k}_o0)\right|
\left\langle\Psi_{J_fM_f}(1 \cdots A) \right|
\; A\, g_{\text{eff}}(0,1)\;
 {\cal A}_{01} \left\{ \left| \chi^{(+)}_\nu ({\bf
k}_i0) \right\rangle \right.  \left. \left| \Psi_{J_iM_i}
 (1\cdots A) \right\rangle \right\} . 
\end{eqnarray}
In this,   distorted wave functions are denoted by $\chi_{\nu}^{\pm}(q)$ 
for an incoming/outgoing proton with spin projection $\nu$,  wave vector 
${\bf k}$, and coordinate set "$q$".   The $A$-nucleon nuclear structure 
wave functions are denoted by $\Psi_{J M}(1\cdots A)$, and  since all of
pairwise interactions between the projectile and   every  target nucleon 
are taken to be the same,  it is convenient to make  cofactor expansions 
i.e.
\begin{equation}
\left| \Psi_{JM}(1,\cdots A) \right\rangle = \frac{1}{\sqrt{A}}
\sum_{j,m} \left| \varphi_{jm}(1) \right\rangle\, a_{jm}(1)\, \left|
\Psi_{JM}(1 \cdots A) \right\rangle\ . \label{cofactor}
\end{equation}
Thus the transition amplitudes expand to the form
\begin{eqnarray}
{\cal T} &=& \sum_{j_1,j_2} \left\langle\Psi_{J_fM_f}(1\cdots A)
\right| a^{\dagger}_{j_2m_2}(1)\ a_{j_1m_1}(1) \left| \Psi_{J_iM_i}(1
\cdots A) \right\rangle
\nonumber\\
&&\hspace*{1.0cm} \times
\left\langle \chi^{(-)}_{\nu^\prime}({\bf k}_o0)\right|
\left\langle \varphi_{j_2m_2}(1) \right| \ g_{\text{eff}}(0,1)\
{\cal A}_{01} \left\{ \left| \chi^{(+)}_\nu ({\bf k}_i0)
\right\rangle
\ \left| \varphi_{j_1m_1} (1) \right\rangle \right\},
\end{eqnarray}
where the many-body  matrix  elements of   particle-hole  operators  are 
expressed by
\begin{eqnarray}
\rho &=& \left\langle\Psi_{J_fM_f}(1,\cdots A) \right|
a^{\dagger}_{j_2m_2}(1)\, a_{j_1m_1}(1) \left| \Psi_{J_iM_i}(1 \cdots
A) \right\rangle
\nonumber\\
&=&  \sum_{I(N)} (-1)^{(j_1 - m_1)} 
\left\langle j_1 j_2 m_1 -m_2 |I -N \right\rangle
\left\langle \Psi_{J_f M_f} |[ a^{\dagger}_{j_2}(1) \times a_{j_1}(1) ]^{I N} |
\Psi_{J_i M_i} \right\rangle
\nonumber\\
&=&  \sum_{I(N)} (-1)^{(j_1 - m_1)}
\left\langle j_1 j_2 m_1 -m_2 |I -N \right\rangle
\ \frac{1}{\sqrt{2J_f +1}} \ 
\left\langle J_i I M_i N |J_f M_f \right\rangle 
S_{j_1 j_2 I}^{J_i \to J_f} ,
\end{eqnarray}
on using the Wigner-Eckart theorem. The OBDME in the above equation  are 
the reduced matrix elements,
\begin{equation}
S_{j_1 j_2 I}^{J_i \to J_f} = \left\langle \Psi_{J_f} ||
[ a^{\dagger}_{j_2}(1) \times a_{j_1}(1) ]^{I } 
||\Psi_{J_i} \right\rangle,
\end{equation}
and  so  carry the  multi-nucleon aspects of nuclear structure tested in
this theory.               Then the transition amplitude can be written, 
\begin{eqnarray}
{\cal T}&=& \sum_{j_1,j_2,m_1,m_2,I(N)}
(-)^{(j_1-m_1)} \frac{1}{\sqrt{2J_f+1}}\,
\left\langle J_i\, I\, M_i\, N \vert J_f\, M_f \right\rangle
\left\langle j_1\, j_2\, m_1\, -m_2 \vert I\, -N \right\rangle
\, S_{j_1\, j_2\, I}^{(J_i \to J_f)}\
\nonumber\\
&&\hspace*{2.0cm} \times
\left\langle \chi^{(-)}_{\nu^\prime}({\bf k}_o0)\Bigr|
\left\langle \varphi_{j_2m_2}(1) \right| \
g_{\text{eff}}(0,1)\
\Bigr| {\cal A}_{01} \left\{ \left| \chi^{(+)}_\nu ({\bf k}_i0)
\right\rangle
\ \left| \varphi_{j_1m_1} (1) \right\rangle \right\}\right\rangle\ .
\end{eqnarray}
As  with  the  generation of the elastic scattering, and so also of  the
distorted wave functions for use in the DWA evaluations, antisymmetry of
the  projectile with the individual  bound nucleons is treated  exactly.  
The associated knock-out (exchange) amplitudes contribute importantly to
the  scattering  cross section, both in magnitude and shape.

\subsubsection{The effective $NN$ interaction}

     A key element in both the $g$-folding and DWA prescriptions is  the
effective $NN$ interaction, $g_{eff}(0,1)$.        This we require to be 
specified in coordinate space and in a form that can be  used  with  the 
DWBA98 programs~\cite{Ra98}.  For that, these effective interactions can 
be constructed having   central ($C$),  tensor ($S_{12}$),  and two-body 
spin-orbit ($L\cdot S$) components; each of which has a form factor that 
is a sum of Yukawas of various ranges.  Each of those Yukawas can have a 
complex strength which is dependent upon  both  the  incident energy and 
the density of the nucleus.  With $r =\vert {\bf r_0} - {\bf r_1} \vert$
and energy $\omega$, 
\begin{equation}
g_{eff}^{ST}(r,\omega) = \sum_{i} <\theta _{i}>
\sum_{j=1}^{n_i} S_j^{(i)}(\omega)
\frac{e^{-\mu^{(i)}_j r}}{r} .
\label{eff-g}
\end{equation}
Here  $\theta_i$  are  the  characteristic operators for central  forces
($i=1$), $\{ 1, (\sigma\cdot\sigma), (\tau\cdot\tau), (\sigma\cdot\sigma
\tau\cdot\tau) \}$, for the tensor force ($i=2$), ${\{ \bf S_{12}} \} $, 
and  the  two-body spin-orbit force ($i=3$),$\{ {\bf L}\cdot{\bf S} \}$.   
$S^{(i)}_{j}(\omega)$  are  complex,  energy-    and    medium-dependent 
strengths, $\mu^{(i)}_{j}$ are the inverse ranges  of  the  interaction, 
and $j$ represents the set of the inverse ranges chosen.   In principle, 
the number of strengths and inverse ranges ($n_i$)  chosen  can  be   as 
large as one likes,  though for  all operators  and for non-relativistic 
energies, $n_i = 4$ seems to be sufficient for one to reproduce the  on-
and half-off-shell $g$-matrices within 32 $NN$ $S, T$ channels.       We
consider what those $g$-matrices are,   and how the $g_{eff}$ is  mapped 
against them, next.

The   nuclear   $g$-matrices   we   take   to   be   solutions   of  the 
Bethe-Brueckner-Goldstone (BBG)  equations  for infinite nuclear matter, 
i.e. of
\begin{equation}
g_{L,L'}^{JST}(p',p;k;k_f) =
V_{L,L'}^{JST}(p',p)+\frac{2}{\pi}\sum_{l}\lim_{\varepsilon
\rightarrow 0} \int_{0}^{\infty} V_{L,l}^{(JST)}(p',q)
[{\cal H}_{\varepsilon}]g_{l,L'}^{(JST)}(q,p;k,k_f)\, q^2 dq.
\end{equation}
The propagator term, ${\cal H}_{\varepsilon}$, is
\begin{equation}
{\cal H}(q,k,K,k_f)=\frac{\bar Q (q,K;k_f)}{\bar E (q,K;k_f) -\bar
E (q,K;k_f)-i\varepsilon }
\end{equation}
in  which  $\bar Q(q,K;k_f)$  is an angle average Pauli operator with an 
average center of mass momentum $K$ and for Fermi momentum $k_F$. A range 
of  Fermi  momenta  spanning free space to 1.5 central nuclear densities
have been considered.         The energies in the propagators of the BBG 
equations include an auxiliary potential $U$, and are defined by
\begin{equation}
\bar{E}(q,K;k_f)= \frac{\hbar^2}{m} ( q^2+K^2 ) + U( | {\bf q} +
{\bf K} |) + U(| {\bf q} - {\bf K} |).
\end{equation}
Full details of these quantities are found in Ref.~\cite{Am00}.

The coordinate space effective $NN$ interactions of    Eq.~(\ref{eff-g}) 
then  have  been  defined by mapping double Bessel transforms of them to 
the on- and a range of  half-off-shell values of those $g$-matrices with 
Fermi momenta set  by  the  density of  the nucleus at the central point 
between the pair. Details are given in Ref.~\cite{Am00}.

\subsection{Models of structure}
\label{structures}

   From the specifics of the scattering potentials and amplitudes  given
above, two details  need be provided by the model chosen to describe the
structure of the target.             Those details are the OBDME and the
single-nucleon bound-state wave functions.    For the nuclei considered,
the models  from  which  those  properties  have  been  determined   are
discussed in brief next.        We consider 10 single-particle states in
scattering calculations,  each  identified by the state number listed in 
Table~\ref{nomenc}.
\begin{table}[ht]
\begin{ruledtabular}
\caption{\label{nomenc} Nomenclature of single-particle orbits.}
\begin{tabular}{cccccc}
ID & $nl_j$ & ID & $nl_j$ & ID & $nl_j$ \\
\hline
 1 &  $0s_{\frac{1}{2}}$ & 4 & $0d_{\frac{5}{2}}$ & 7 & $0f_{\frac{7}{2}}$ \\
 2 &  $0p_{\frac{3}{2}}$ & 5 & $0d_{\frac{3}{2}}$ & 8 & $0f_{\frac{5}{2}}$ \\
 3 &  $0p_{\frac{1}{2}}$ & 6 & $1s_{\frac{1}{2}}$ & 9 & $1p_{\frac{3}{2}}$ \\
   & & & & 10 & $1p_{\frac{1}{2}}$ \\
\end{tabular}
\end{ruledtabular}
\end{table}

\subsubsection{No-core shell model for ${}^{12}$C}

A no-core  shell model calculation has been made to define the  spectrum
of ${}^{12}$C, the ground-state shell occupancies, and the OBDME for the
excitation of the $2^+_1$, (4.43 MeV) state~\cite{Ka96}.        The code
OXBASH~\cite{Ox86} was used with augmented MK3W potentials  and  with  a
complete $(0+2)\hbar\omega$ basis to define  the  positive parity states
in the spectrum to 20 MeV excitation.     With the exception of the well
known strongly deformed $0^+_2$ state,       all other states were found
within a few hundred keV of their known excitations.  In particular, the
predicted excitation energy of the $2^+_1$ state was 4.62 MeV.  

The  ground-state   occupancies  found  from this shell model study  are 
dominantly those of the $0s_{\frac{1}{2}}$ (1.964)-, $0p_{\frac{3}{2}}$
(3.054)-, and $0p_{\frac{1}{2}}$ (0.842)-shells    (for both protons and
neutrons); the remaining 0.14 nucleons being in the higher orbits.   The
OBDME for inelastic scattering are given in Table~\ref{C-obdme}.
\begin{table}[ht]
\begin{ruledtabular}
\caption{\label{C-obdme}
The OBDME for excitation of the $2^+_1$ state in  ${}^{12}$C.}
\begin{tabular}{ccccccccc}
$j_2$ & $j_1$ & $S_{j_1 j_2,2}$  & $j_2$ & $j_1$ & $S_{j_1 j_2,2}$  &
$j_2$ & $j_1$ & $S_{j_1 j_2,2}$  \\
\hline
 2 & 2 &    0.5609 & 7 & 2 & $-$0.1391 &10 & 8 & $-$0.0004 \\
 3 & 2 & $-$1.0706 & 8 & 2 &    0.0530 & 2 & 9 &    0.0093 \\  
 2 & 3 &    0.7728 & 9 & 2 & $-$0.0136 & 3 & 9 & $-$0.0042 \\
 4 & 1 & $-$0.1586 &10 & 2 &    0.0270 & 7 & 9 & $-$0.0011 \\ 
 5 & 1 &    0.1356 & 8 & 3 & $-$0.0526 & 8 & 9 &    0.0007 \\
 1 & 4 & $-$0.1586 & 9 & 3 &    0.0003 & 2 &10 &    0.0055 \\  
 4 & 4 &    0.0174 & 2 & 7 & $-$0.0576 & 8 &10 & $-$0.0004 \\
 5 & 4 & $-$0.006  & 7 & 7 & $-$0.0038 & & & \\
 6 & 4 &    0.0026 & 8 & 7 &    0.0010 & & & \\
 1 & 5 & $-$0.0949 & 9 & 7 & $-$0.0008 & & & \\
 4 & 5 &    0.0042 & 2 & 8 & $-$0.0398 & & & \\
 5 & 5 &    0.0087 & 3 & 8 & $-$0.0071 & & & \\
 6 & 5 &    0.0012 & 7 & 8 & $-$0.0020 & & & \\
 4 & 6 &    0.0013 & 8 & 8 & $-$0.0007 & & & \\
 5 & 6 & $-$0.0016 & 9 & 8 & $-$0.0001 & & & \\
\end{tabular}
\end{ruledtabular}
\end{table}
Using these OBDME in evaluations of the $B(E2; 2^+ \to 0^+ (gs))$   with
bare charges gave a value of 6.26 $e^2$-fm$^4$ when  HO  wave  functions
with oscillator length  of 1.7 fm was used. This compares very favorably
with the known value of 7.77$e^2$-fm$^4$, especially when the  Cohen and
Kurath ($0\hbar\omega$ shell model) gives 3.26 $e^2$-fm$^4$,   on which,
the no-core large space model of Navratil, Vary, and Barrett~\cite{Na00}
only slightly improves.  By holding fast to an  $NN$ interaction defined
for free space $NN$ collisions in their model,   and as they note in the
article,  they  did  not   account   sufficiently   for   multi-particle
correlations in nuclear structure.

\subsubsection{
PHF and SHF plus shell models for ${}^{20}$Ne, ${}^{24}$Mg, ${}^{28}$Si,
and ${}^{40}$Ca}

Using the SHF  model of structure, charge-density distributions and  the 
associated nuclear radii have been   calculated  previously~\cite{Ri03}.
The  resulting  wave functions  gave form factors in very good agreement
with available data from electron scattering.    Two forms of the Skyrme
interaction were used,       the so-called SkX$_{csb}$~\cite{Br00}   and
SkM*~\cite{Ba82} interactions.  The SkX$_{csb}$ Hamiltonian is based  on
the SkX Hamiltonian~\cite{Br98} with a    charge-symmetry-breaking (CSB)
interaction   added   to  account  for  nuclear   displacement  energies
\cite{Br00}.   The charge densities from all three calculations are very
similar and so we only use those  determined from  the SkX$_{csb}$ model
(referred to hereafter as simply ``SkX'').          There are some small
$\le 5\%$ differences in the interior densities  found with these models
but they have little effect on   scattering   results; especially of the
total reaction cross sections~\cite{Am06}.      Generally, with this SHF
method, good agreement between theory and experiment has   been achieved
in    extensive   comparisons  of   measured    nuclear   charge-density
distributions  with  calculated values for  $p$-shell,  $sd$-shell,  and
$pf$-shell nuclei and some  selected  magic and semi-magic nuclei  up to 
${}^{208}$Pb.      With the pure SkX model, proton and neutron densities
differ slightly. While such differences do not effect scattering results
much, the small differences have been noted in determining a   value for
the neutron skin in ${}^{208}$Pb~\cite{Ka02}. 

The ground-state shell   occupancies found from the SHF plus shell model
studies~\cite{Ri03} of the nuclei, ${}^{20}$Ne to ${}^{40}$Ca are listed
in Table~\ref{shell-occs}.
\begin{table}[ht]
\begin{ruledtabular}
\caption{\label{shell-occs}
Shell occupancies from the SHF plus shell model calculations
of the listed nuclei}
\begin{tabular}{ccccc}
ID  & ${}^{20}$Ne & ${}^{24}$Mg & ${}^{28}$Si & ${}^{40}$Ca\\
\hline
1 &  2.000 & 2.000 & 2.000 &  2.000 \\
2 &  4.000 & 4.000 & 4.000 &  4.000 \\ 
3 &  2.000 & 2.000 & 2.000 &  2.000 \\
4 &  1.209 & 2.990 & 4.623 &  6.000 \\
5 &  0.283 & 0.563 & 0.673 &  3.090 \\
6 &  0.508 & 0.448 & 0.704 &  1.800 \\
7 &   ---  & ---   & ---   &  0.990 \\
8 &   ---  & ---   & ---   &  0.000 \\
9 &   ---  & ---   & ---   &  0.120 \\
10 &  ---  & ---   & ---   &  0.000 \\
\end{tabular}
\end{ruledtabular}
\end{table}

While  much  success  has  been  had  using  the  SHF densities and wave
functions generated using the SkX model,  the  canonical  wave functions
may not have a desirable long range character.      So we have also used
Woods-Saxon (WS) single-nucleon bound-state wave functions  to   compare
scattering results.     For these nuclei, the SkX calculations only give
ground-state properties while SM studies,     that do provide transition
OBDME~\cite{Ka95},         have not been made with a large enough basis.
Consequently, we resorted to PHF evaluations to define the OBDME   to be
used in calculations of inelastic scattering to the $2^+_1$ states.  For
completeness, these OBDME are listed in Table~\ref{obdmes}.
\begin{table}[ht]
\begin{ruledtabular}
\caption{\label{obdmes}
OBDME for the transitions to the $2^+_1$ states (values for which at
least one entry $> \pm 0.02$).
$j_2, (j_1)$ are a list of particle (hole) terms with identification as
in Table~\ref{shell-occs}.}
\begin{tabular}{ccccc} 
$j_2$ & $j_1$ & ${}^{20}$Ne  & ${}^{24}$Mg  & ${}^{28}$Si \\
\hline
    4 & 4 &    0.592  &    0.8308 &    0.7088 \\
    5 & 4 & $-$0.169  & $-$0.5738 & $-$0.6542 \\
    6 & 4 &    0.601  &    0.6818 &    0.6331 \\
    4 & 5 &    0.121  &    0.5954 &  0.6111   \\
    5 & 5 &    0.107  & $-$0.0151 &    0.4549 \\
    6 & 5 &    0.193  &    0.1969 &    0.1481 \\
    4 & 6 &    0.752  &    0.5428 &    0.6782 \\
    5 & 6 & $-$0.341  & $-$0.2321 & $-$0.2865 \\
    4 & 1 & $-$0.190  & $-$0.1641 & $-$0.2475 \\
    5 & 1 &    0.128  &    0.1229 &    0.1541 \\
    1 & 4 & $-$0.148  & $-$0.2092 & $-$0.2250 \\  
    1 & 5 & $-$0.081  & $-$0.1027 & $-$0.0950 \\
    7 & 2 & $-$0.147  & $-$0.2270 & $-$0.2074 \\
    8 & 2 &    0.089  &    0.1054 &    0.1001 \\
    9 & 2 & $-$0.058  & $-$0.0368 & $-$0.0393 \\
   10 & 2 &    0.049  &    0.0302 &    0.0287 \\ 
    8 & 3 &    0.180  & $-$0.2169 & $-$0.1864 \\
    9 & 3 & $-$0.060  & $-$0.0377 & $-$0.0318 \\
    2 & 7 & $-$0.131  & $-$0.1762 & $-$0.1659 \\
    9 & 7 & $-$0.027  & $-$0.0464 & $-$0.0456 \\
    2 & 8 & $-$0.067  & $-$0.0819 & $-$0.0802 \\
    3 & 8 & $-$0.135  & $-$0.1685 & $-$0.1492 \\
    7 & 8 &    0.004  &    0.0046 & $-$0.0044 \\
    9 & 8 & $-$0.014  & $-$0.0216 & $-$0.0217 \\
   10 & 8 & $-$0.018  & $-$0.0368 & $-$0.0346 \\
    2 & 9 & $-$0.042  & $-$0.0274 & $-$0.0303 \\
    3 & 9 &    0.016  &    0.0283 &    0.0246 \\
    7 & 9 & $-$0.035  & $-$0.0602 & $-$0.0565 \\
    8 & 9 &    0.018  &    0.0280 &    0.0270 \\
    9 & 9 & $-$0.022  & $-$0.0178 & $-$0.0208 \\
    2 &10 & $-$0.036  & $-$0.0226 & $-$0.0221 \\
    8 &10 & $-$0.024  & $-$0.0476 & $-$0.0430 \\
    9 &10 & $-$0.016  & $-$0.0147 & $-$0.0148 \\
\end{tabular}
\end{ruledtabular}
\end{table}


\section{ Results and discussions}
\label{Results}

We analyze the differential cross sections and  analyzing  powers   from
elastic and inelastic  (to $2^+_1$ states)  scattering  off  ${}^{12}$C,
${}^{20}$Ne, ${}^{24}$Mg, ${}^{28}$Si and ${}^{40}$Ca. Analyses are made
at non-relativistic energies for each case where data  exist for  proton
scattering.

\subsection{Elastic and inelastic scattering of protons from ${}^{12}$C}

\begin{figure}[h]
\scalebox{0.60}{\includegraphics*{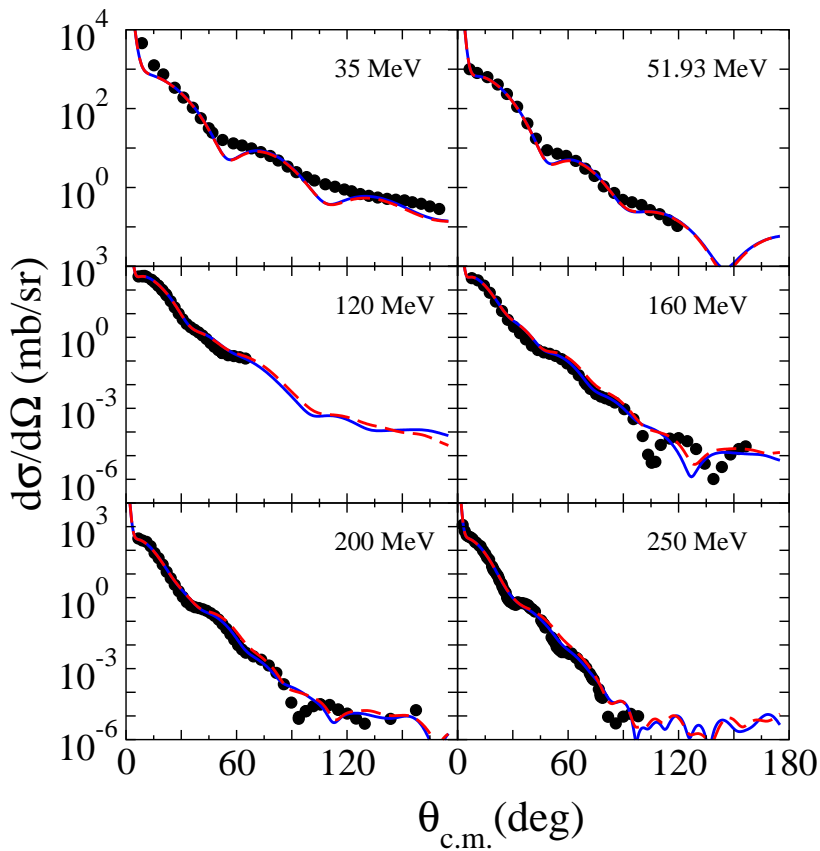}}
\caption{\label{Fig1}(Color online) $g$-folding model
predictions of cross sections for the elastic scattering of 35,
51.93, 120, 160, 200 and 250 MeV protons from ${}^{12}$C compared
with data~\cite{Pi86,Oh80,Co81,Me83,Me81,Me88}. The
predictions found by using HO wave functions are shown by the solid
curves while those obtained with the WS single-particle wave
functions are displayed by the dashed curves. }
\end{figure}

In Fig.~\ref{Fig1}, the cross sections and analyzing powers for  elastic
scattering data~\cite{Pi86,Oh80,Co81,Me83,Me81,Me88}  of 35, 51.93, 120,
160, 200 and 250 MeV protons from ${}^{12}$C  are  compared  with    the
results found using the $g$-folding model. 
Traditionally  one chooses either HO or WS functions for the bound-state 
single-particle wave functions.      The solid curves in this figure are 
the results obtained using the HO model of structure,   while the dashed
curves are those obtained when WS bound-state wave functions are used. As
shown in Fig.~\ref{Fig1}, the predictions of cross sections are in quite
good agreement with experimental data up to $170^\circ$ scattering,  
in those cases where the data extend to this scattering angle.
But the minima of 35 and 51.93 MeV  calculated  results   are  more  sharply
defined than seen in data.     It is evident that back angle data are
not well described by the predictions,   but the cross section values are very
small, usually much less than a mb/sr.   Small variations in the details 
used in our calculations,  as well as other reaction process effects not
considered within the $g$-folding approach  (if  such  have  influence), 
will affect small cross-section values most obviously.

\begin{figure}[h]
\scalebox{0.60}{\includegraphics*{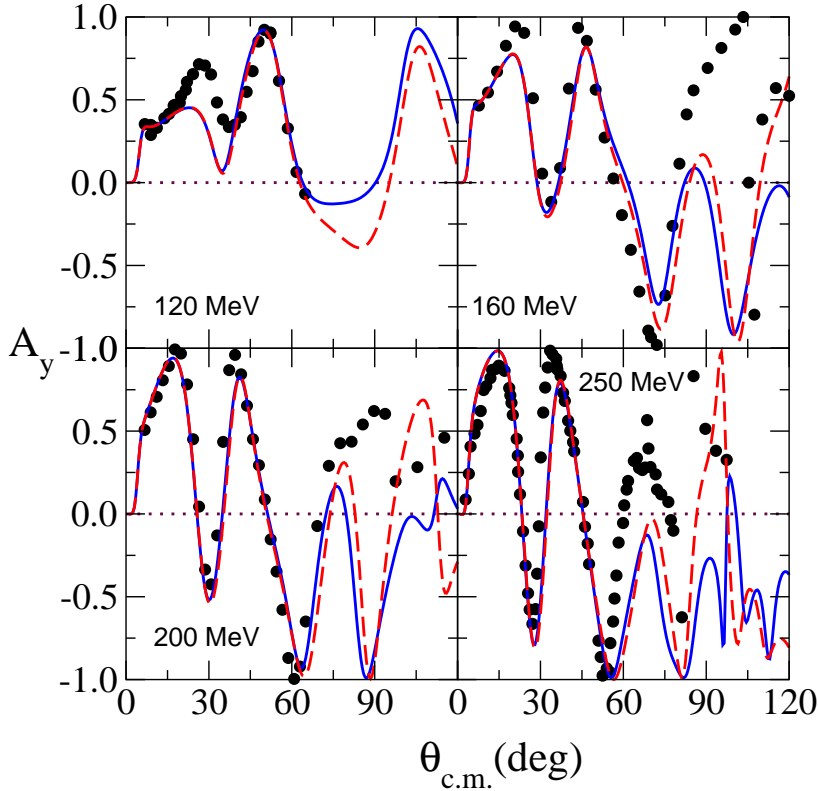}}
\caption{\label{Fig2}(Color online) $g$-folding model
predictions of analyzing powers for the elastic scattering of 120,
160, 200 and 250 MeV protons from ${}^{12}$C compared with data
~\cite{Co81, Me83, Me81, Me88}.  The notation is as used in
Fig.~\ref{Fig1}.}
\end{figure}
In Fig.~\ref{Fig2},  we compare our calculated analyzing powers from
proton elastic scattering on ${}^{12}$C with the data that  have been
taken at  120,  160, 200 and 250
MeV~\cite{Co81,Me83,Me81,Me88}.            Clearly our predictions match
observation quite well up to $\sim 70^\circ$ scattering and for all four
energies.  But there are noticeable discrepancies at large angles.    HO
and  WS  results  are shown by the solid and dashed curves respectively, 
and  they  are  almost  indistinguishable  over  the range of scattering 
angles at which we find good representation of the data.     Differences
between the HO and WS results appear explicitly as the angle  increases,
though neither result makes a match to observation there.   But one must
remember that analyzing powers are normalized against     the scattering
cross section;  data  against  measured  values and  theoretical ones
against theoretical cross sections. Thus wherever theory does not give a
sufficiently
good representation of cross-section data,    a match to analyzing power
data can only be considered fortuitous.   Equally a mismatch must not be 
taken necessarily as consequential.

\begin{figure}[h]
\scalebox{0.60}{\includegraphics*{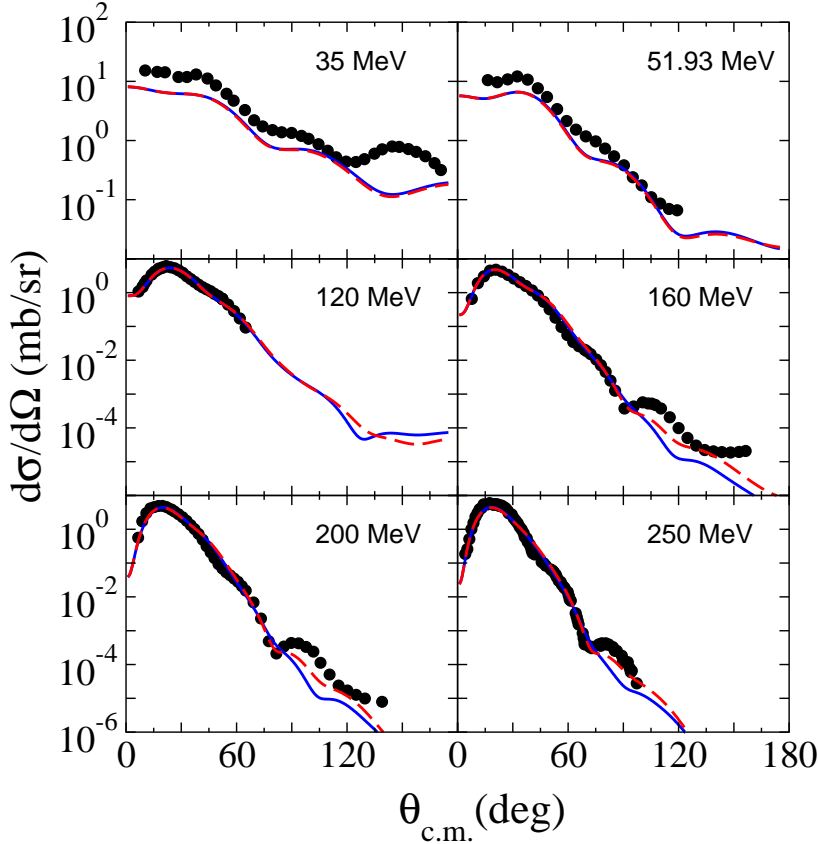}}
\caption{\label{Fig3}(Color online) DWA cross sections for the
inelastic scattering of 35, 51.93, 120, 160, 200 and 250 MeV protons
from the excitation of the $2^+$;4.44 MeV state in ${}^{12}$C
compared with data ~\cite{Pi86, Oh80, Co81, Hu83, Hi88}. The
notation is as used in Fig.~\ref{Fig1}. }
\end{figure}

The cross sections from inelastic scattering of 35, 51.93, 120, 160, 200
and 250 MeV protons exciting the $2^+$; 4.44 MeV state in ${}^{12}$C are
displayed in Fig.~\ref{Fig3}. 
The results of our DWA calculations reproduce the shape and magnitude of
these cross sections for all but the lowest energies, and to quite small
magnitudes.  Again the differences between results found using HO and WS
single-nucleon bound-state wave functions are negligible until   one has
very small cross-section values. The HO and WS results are displayed  by
the solid and dashed curves respectively.       These results bespeak of
appropriate structure since the no-core shell model structure  not  only 
gave good  results
for the $B(E2)$ and electron scattering form factors~\cite{Ka95},    but
now also for cross sections formed using the $g$-folding approach   with 
the effective $NN$ interactions for many different energies.    There is
clearly some other process required to explain the data  at 35 and 51.93
MeV.                           Our $g$-folding results underestimate the 
data~\cite{Pi86,Oh80,Co81,Hu83,Hi88}   by   factors   of   2.0  and  1.5 
respectively. Further there is a noticeable difference in shape  between
our results and the data at large scattering angles for the 35 MeV case.
Such is not caused by an inadequacy in the effective $NN$ interaction. 
Recently, good results for these and even lower energies have been found
in studies of ${}^6$He scattering from hydrogen,        both elastic and
inelastic exciting the $2^+_1$ state of ${}^6$He.        We believe that 
there are competing scattering processes that have been ignored;     and
processes that relate to this target and for these energies.     Virtual
excitation of giant resonances seems a most likely cause.   Past studies
of inelastic scattering of protons from ${}^{12}$C have indicated  that,
at 35 MeV specifically,         contributions from virtual excitation of
isoscalar $E2$  and $E3$ resonances can contribute noticeably~\cite{Ge75}.
They do so as specific second order processes,        first by enhancing
direct scattering amplitudes (in which the incoming   proton is also the
emergent one) and by exchange amplitude contributions   in   which   the 
incident proton is trapped with the giant resonance formed  subsequently
decaying by emitting the detected proton. That exchange process leads to
cross-section contributions that are symmetric  about  90$^\circ$ and of 
about a mb/sr in size~\cite{Ge75}.    The correction at large scattering
angles needed in the 35 MeV result is  characteristic of the corrections
just such an exchange process would  give~\cite{Ge75},         while the
enhancements needed for both the 35 and  52 MeV results   are consistent 
also with contributions from  the  direct  effect  of  the  second-order
process involving giant resonances.

\begin{figure}[h]
\scalebox{0.60}{\includegraphics*{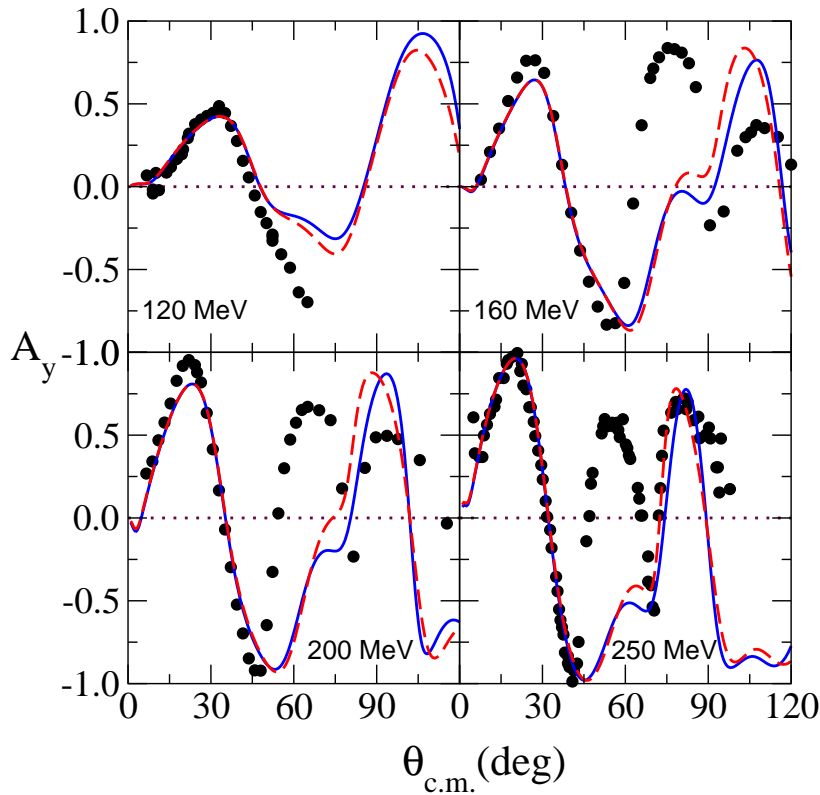}}
\caption{\label{Fig4}(Color online) Analyzing powers for the
inelastic scattering of 120, 160, 200 and 250 MeV protons from the
excitation of the $2^+$; 4.44 MeV state in ${}^{12}$C compared with
data ~\cite{Co81, Hu83, Hi88}. 
}
\end{figure}

The  DWA  results for the analyzing power from 120, 160, 200 and 250 MeV 
protons exciting the  $2^+$; 4.44 MeV  state in ${}^{12}$C are compared 
with
the data ~\cite{Co81,Hu83,Hi88} in Fig.~\ref{Fig4}.    The results found
using  HO  and  WS  wave functions again are displayed by the solid  and 
dashed curves respectively.    Those results have similar structures and 
are in good agreement with the data up to $\sim 45^\circ$.     At larger
scattering angles there are noticeable discrepancies between  calculated
results and the data.   For these four energies, by $45^\circ$ the cross
sections are quite small ($\le 1$ mb/sr) and so sensitive observables such 
as the analyzing power,   as they are normalized against cross sections, 
can then show large effects caused by minor problems in detail.

\subsection{Elastic and inelastic scattering of protons from 
${}^{20}$Ne}

\begin{figure}[h]
\scalebox{0.60}{\includegraphics*{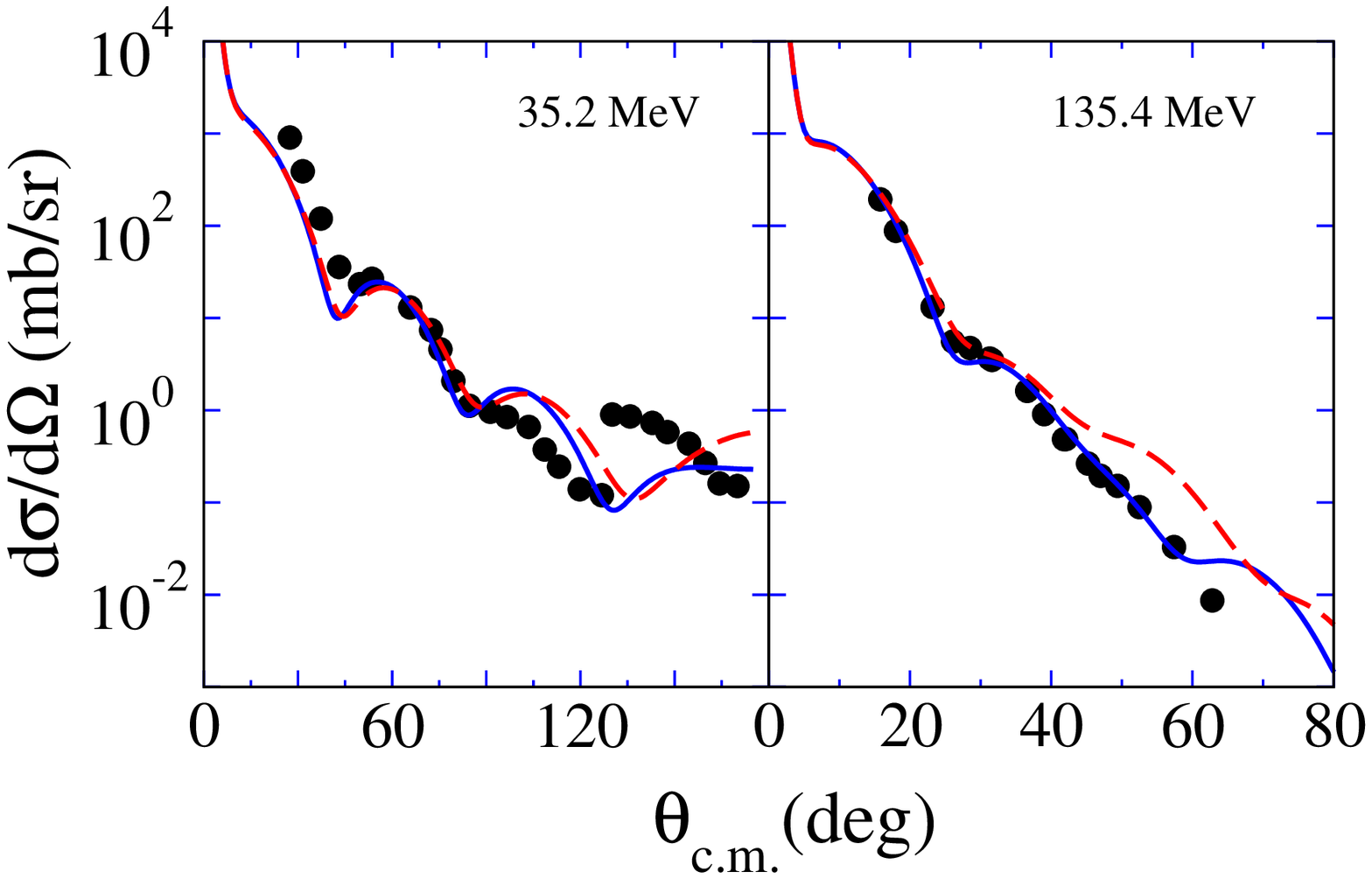}}
\caption{\label{Fig5}(Color online) $g$-folding model
predictions of cross sections for the elastic scattering of 35.2 and
135.4 MeV protons from ${}^{20}$Ne compared with data ~\cite{Co78,
Mu91}.}
\end{figure}

Our $g$-folding model predictions of the cross sections for the  elastic
scattering of 35.2 and 135.4 MeV protons from ${}^{20}$Ne are   compared 
with data in Fig.~\ref{Fig5}.
\begin{figure}[h]
\scalebox{0.60}{\includegraphics*{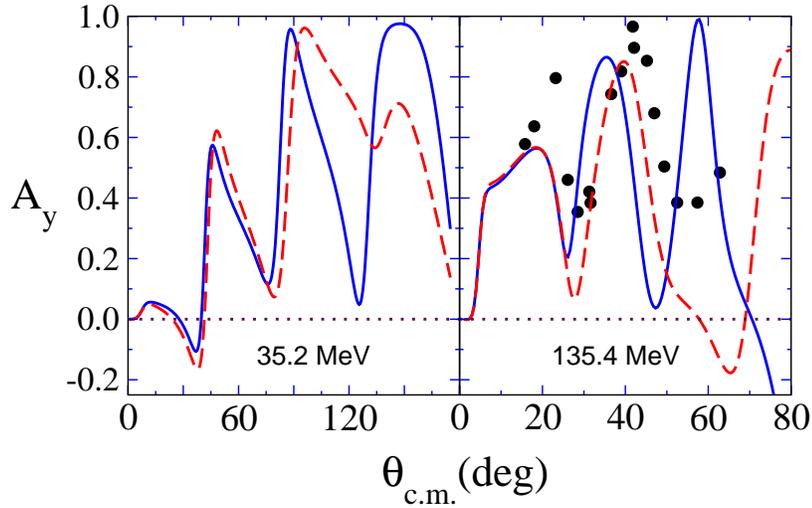}}
\caption{\label{Fig6}(Color online) $g$-folding model
predictions of analyzing powers for the elastic scattering of 35.2
and 135.4 MeV protons from ${}^{20}$Ne compared with data~\cite{Mu91}.}
\end{figure}
Single-particle bound-state wave functions were assumed to be either  a
WS set or those generated with the SHF method (with the SkX interaction).
The solid curves display results obtained using the SHF wave  functions,
while dashed curves show those found with the WS wave functions.     The
135.4 MeV cross-section   data~\cite{Mu91}  are  well  reproduced by our 
calculations, especially when the SHF bound states are used.    However,
that degree of matching may be only fortuitous.  The cross section found
with  35.2  MeV  protons scattering  from  ${}^{20}$Ne  is reasonable to 
$\sim 90^\circ$, by which time the cross-section magnitude is   1 mb/sr,
and  the  mismatch  for  larger  scattering  angles is reminiscent of an 
effect of virtual excitation of a giant resonance.

The elastic scattering analyzing power data ~\cite{Mu91} and $g$-folding
model predictions are displayed in Fig.~\ref{Fig6}.  The notation is the
same as used in Fig.~\ref{Fig5}. There are no available data at 35.2 MeV
and we display those results  only to note  what  differences  there are 
between use of  WS and SHF single-nucleon bound-state wave functions. To
$\sim 90^\circ$ scattering  where both sets of wave functions give  good
cross-section results,         there is practically no difference in the 
analyzing powers found.     They do differ somewhat at larger scattering
angles but then the actual cross sections are small and differ from both
theoretical predictions.  In the case of 135.4 MeV scattering, there are
noticeable differences between predictions and data.    Nevertheless the
$g$-folding model results do show the data trend.

\begin{figure}[h]
\scalebox{0.60}{\includegraphics*{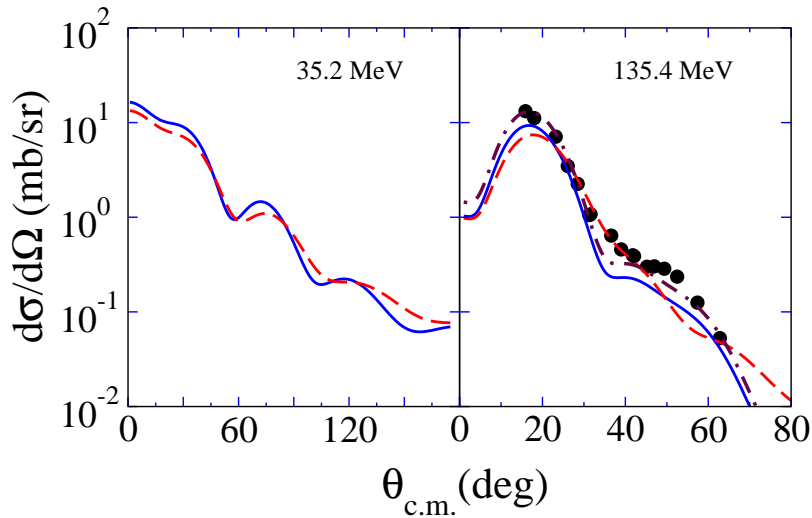}}
\caption{\label{Fig7}(Color online) DWA cross sections for
the inelastic scattering of 35.2 and 135.4 MeV protons from the
excitation of the $2^+$;1.633 MeV state in ${}^{20}$Ne compared with
data ~\cite{Mu91}.}
\end{figure}

Differential cross sections from the inelastic scattering of protons  to
the $2^+$ (1.633 MeV) state in ${}^{20}$Ne found from DWA  calculations
are displayed in Fig.~\ref{Fig7}.    No data have been taken at 35.2 MeV 
incident energy and the predictions made using SHF (solid curve) and  WS
(dashed curve) have been made simply to show that there are but    minor
effects due to the precise nature of the wave functions used.
In the case of 135.4 MeV, DWA results under-predict data~\cite{Mu91}. If
the SHF result is enhanced   by  40\%, a very good fit is found to that 
data.      The dot-dash curve is the SHF cross section multiplied   by a 
factor of 1.4. This enhancement, taken as  a problem of theory, then can 
only be an effect of missing configuration mixing in the (PHF) structure
model.

\begin{figure}[h]
\scalebox{0.60}{\includegraphics*{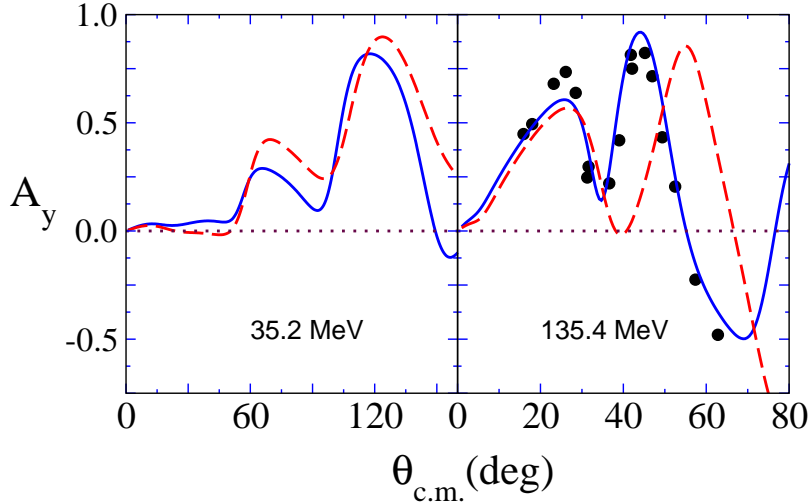}}
\caption{\label{Fig8}(Color online) Analyzing powers for the
inelastic scattering of 35.2 and 135.4 MeV protons from the
excitation of the $2^+$;1.633 MeV state in ${}^{20}$Ne compared with
data ~\cite{Mu91}.}
\end{figure}

The analyzing powers from excitation of the $2^+$ state is displayed  in
Fig.~\ref{Fig8} where, again, the solid and dashed curves depict the SHF 
and WS results respectively. As with the cross sections, the results for
the scattering of 35.2 MeV protons are very similar,     while there are 
more differences between them seen in the 135.4 MeV results. As with the
cross sections, agreement between the SHF result and the 135.4 MeV  
data~\cite{Mu91} for scattering to the $2^+$ state is quite good,  while
the WS result only follows the trend of data.  

In the elastic and inelastic scattering analyzing powers from protons on
${}^{20}$Ne,        the WS results universally are shifted toward higher
momentum transfer values when compared with the SHF results. This is the
prime feature that distinguishes the  choice taken for   single-particle
wave functions.

\subsection{Elastic and inelastic scattering of protons from 
${}^{24}$Mg}

\begin{figure}[h]
\scalebox{0.55}{\includegraphics*{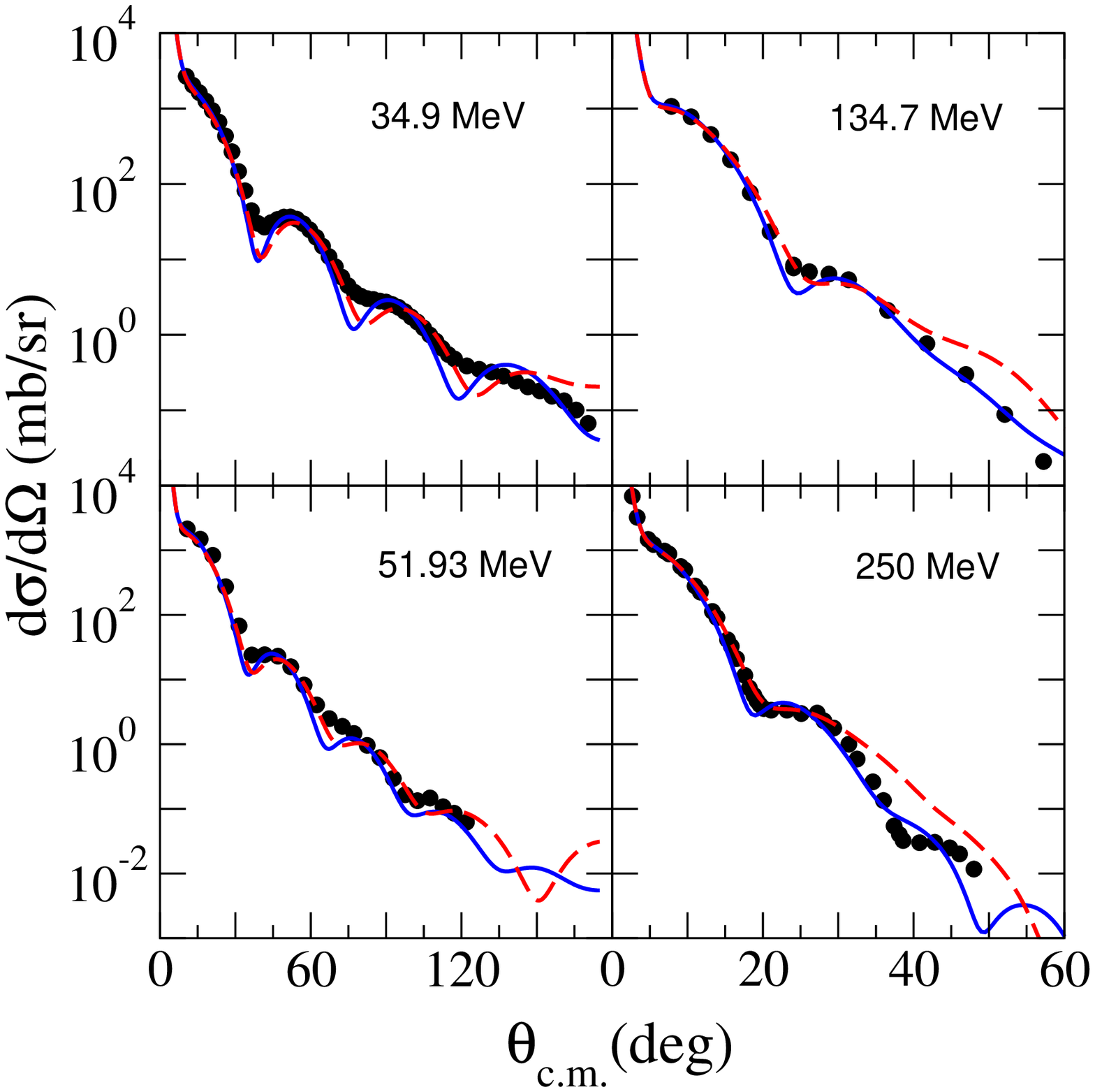}}
\caption{\label{Fig9}(Color online) $g$-folding model
predictions of cross sections for the elastic scattering of 34.9,
51.93, 134.7 and 250 MeV protons from ${}^{24}$Mg compared with 
data~\cite{Ha83, Oh80, Sc82, Hi88}.}
\end{figure}

In Fig.~\ref{Fig9},  differential  cross  sections  for  the     elastic 
scattering of protons from ${}^{24}$Mg with energies of     34.9, 51.93,
134.7, and 250 MeV are shown.    The results of calculations made using 
SHF wave functions are depicted by the solid curves,   while those found
from using WS wave functions are shown by the dashed curves.         SHF
predictions  of  the  elastic  cross  sections made with the $g$-folding 
model agree very well with  the data~\cite{Ha83,Oh80,Sc82,Hi88}   except
for the overly sharp defined minima  they  give with the lower energies.
For the two larger energies, these SHF results give better fits to   the
data than do those found using the WS functions,     noticeably at large
scattering angles.

\begin{figure}[h]
\scalebox{0.55}{\includegraphics*{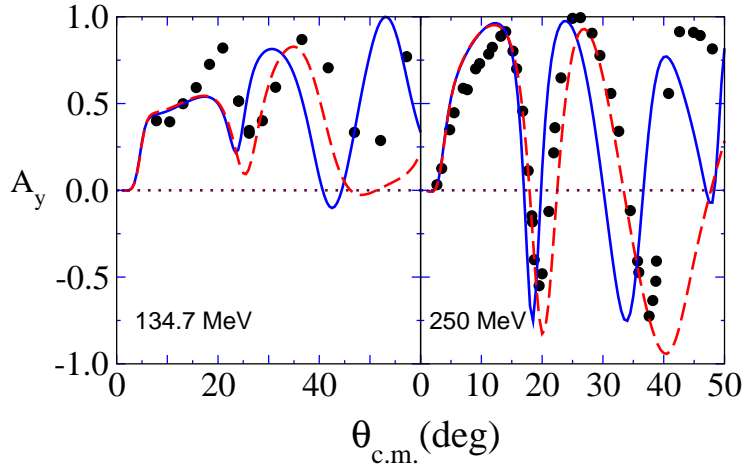}}
\caption{\label{Fig10}(Color online) $g$-folding model predictions of 
analyzing powers for the elastic scattering of 134.7 and 250 MeV protons
from ${}^{24}$Mg compared with data~\cite{Sc82,Hi88}.}
\end{figure}

Analyzing  powers  for  proton  elastic scattering from ${}^{24}$Mg have 
been measured~\cite{Sc82, Hi88} for incident energies of 134.7 and   250
MeV.        That data are compared with our $g$-folding model results in
Fig.~\ref{Fig10}. Again the solid and dashed curves depict results found
using SHF and WS wave functions respectively. There are quite noticeable
differences between these results. In these cases, the WS functions give
the best representation of the data.    As with the ${}^{20}$Ne results,
the WS results are shifted to higher momentum transfer values   compared 
with those of   SHF results. 

\begin{figure}[h]
\scalebox{0.55}{\includegraphics*{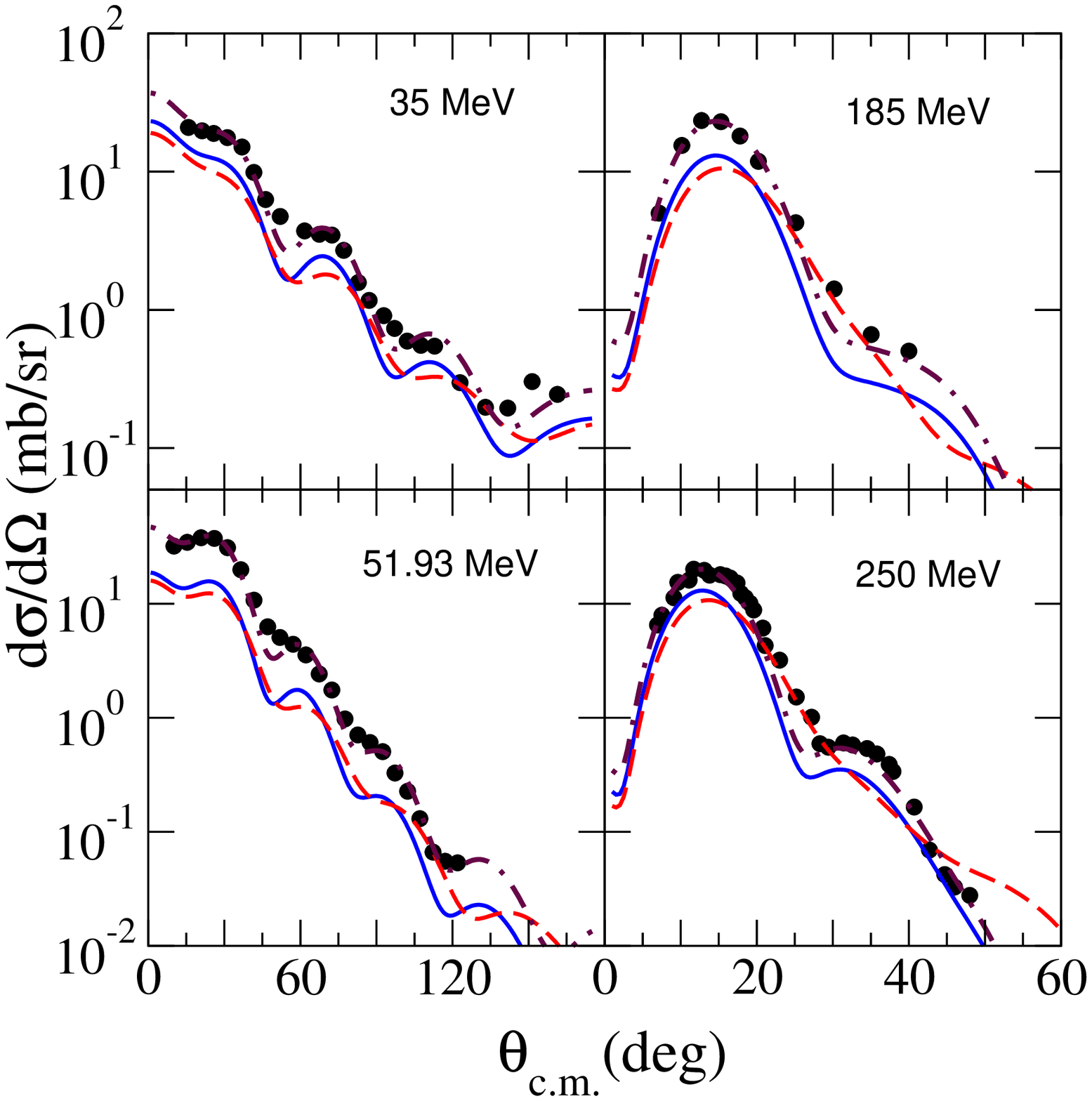}}
\caption{\label{Fig11}(Color online) DWA cross sections for
the inelastic scattering of 35, 51.93, 185 and 250 MeV protons from
the excitation of the $2^+$;1.37 MeV state in ${}^{24}$Mg compared
with data~\cite{Pi86,Oh80,Da67,Hi88}.} 
\end{figure}

DWA  calculated  cross  sections for  35, 51.93, 185 and 250 MeV  proton
inelastic  scattering  to  the  $2^+_1$ state  at 1.37 MeV excitation in 
${}^{24}$Mg   are   compared  with    data~\cite{Pi86,Oh80,Da67,Hi88} in 
Fig.~\ref{Fig11}.    The results that have been obtained using SkX model 
wave functions are displayed by the solid curves.      The dashed curves
portray those found by using the WS set of wave functions.     The third
result in each panel depicted by the dot-dashed curve is the   SkX model
cross section but enhanced by scale factors of 1.6, 2.5, 1.77 and   1.54
for the 35, 51.93, 185 and 250 MeV cases respectively.      These results
compare very well with the data.
\begin{figure}[h]
\scalebox{0.55}{\includegraphics*{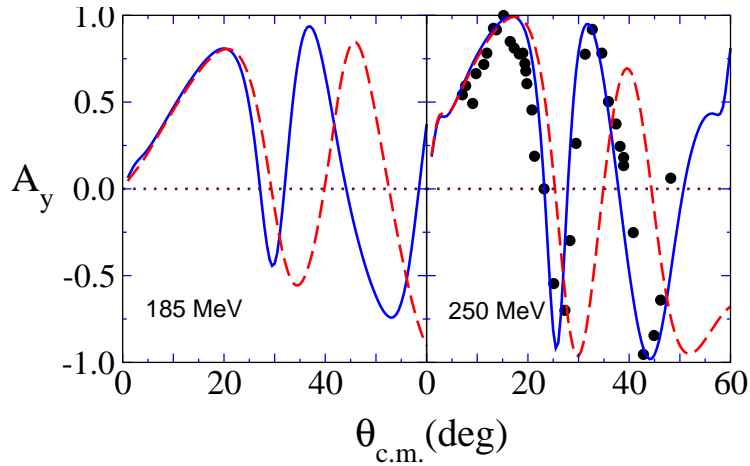}}
\caption{\label{Fig12}(Color online) Analyzing powers for the
inelastic scattering of 185 and 250 MeV protons from the excitation
of the $2^+$;1.37 MeV state in ${}^{24}$Mg compared with data
~\cite{Hi88}.}
\end{figure}

Analyzing  power  data for inelastic scattering to the $2^+_1$ state  in 
${}^{24}$Mg have been taken with 250 MeV protons~\cite{Hi88}. We compare
that data with the results of our calculations that were  made using the
SHF and WS sets of single-particle bound-state wave functions  (notation
as with Fig.~\ref{Fig11}) on the right of Fig.~\ref{Fig12}.  The result
obtained using the SkX model is in quite good  agreement with the data; 
distinctly better than the result found with the WS functions.    As with
other results, the analyzing power predicted using the  WS  functions is
shifted to larger angular momentum transfer in comparison with the   SkX
model one. That distinction is also seen with the predictions   shown in
Fig.~\ref{Fig12} for an energy of 185 MeV.

\subsection{Elastic and inelastic scattering of protons from ${}^{28}$Si}

\begin{figure}
\scalebox{0.5}{\includegraphics*{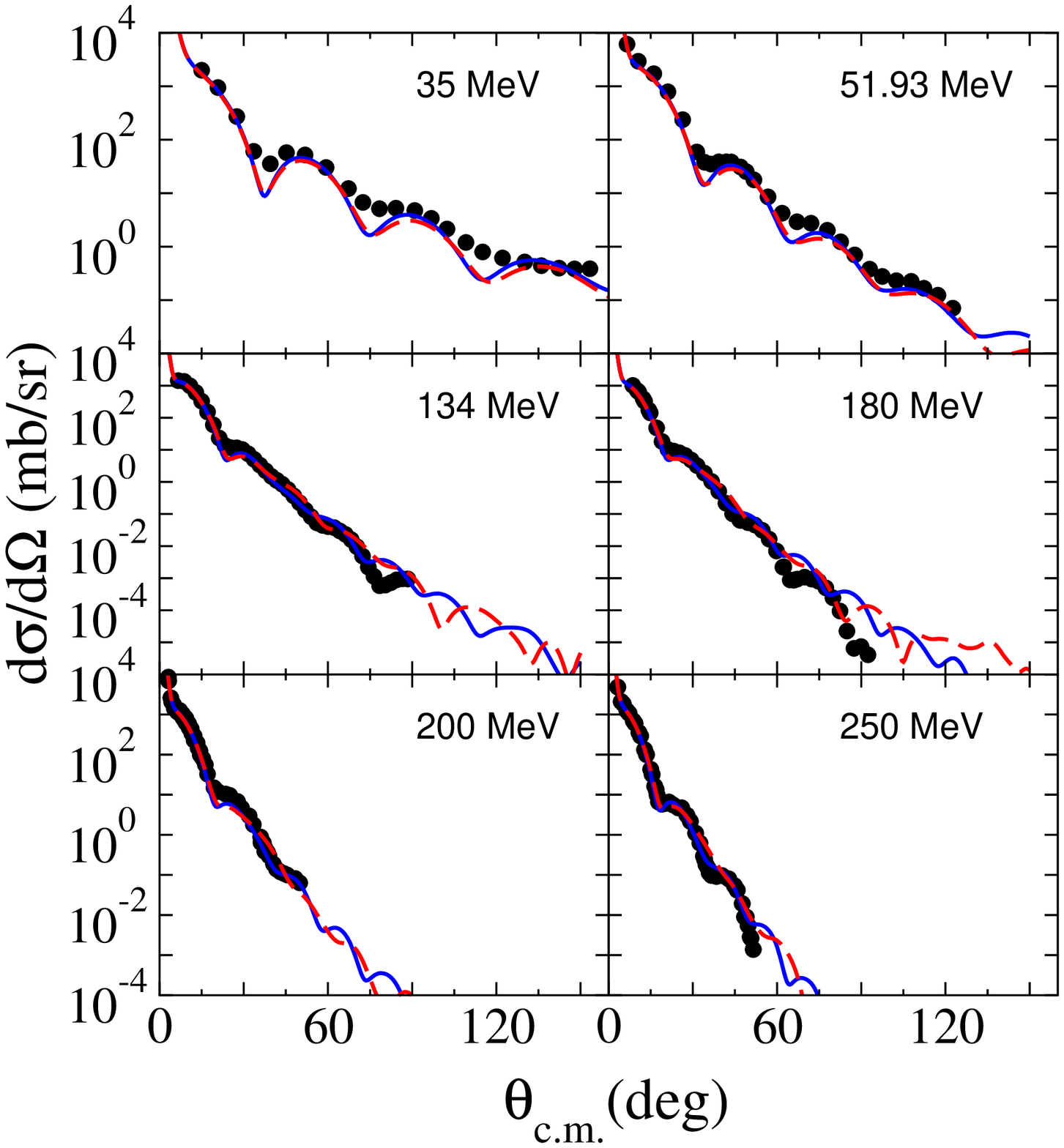}}
\caption{\label{Fig13}(Color online) $g$-folding model
predictions of cross sections for the elastic scattering of 35,
51.93, 134, 180, 200 and 250 MeV protons from ${}^{28}$Si compared
with data~\cite{Pi86,Oh80,Ol84,Hi88}.}
\end{figure}

The differential cross sections for 35, 51.93, 134, 180, 200 and 250 MeV
protons   elastically   scattered   from   ${}^{28}$Si   are  shown   in
Fig.~\ref{Fig13}. Therein data~\cite{Pi86,Oh80,Ol84,Hi88}  are  compared
with the results  of  calculations  made  using  the  SHF and WS sets of 
single-nucleon bound-state wave functions for this nucleus.  The SHF and
WS results are displayed by the solid and dashed
curves, respectively.             At 35 and 51.93 MeV, the minima of the 
$g$-folding results are    more pronounced than as seen in the data. The
magnitudes of the cross-section predictions also 
\begin{figure}
\scalebox{0.50}{\includegraphics*{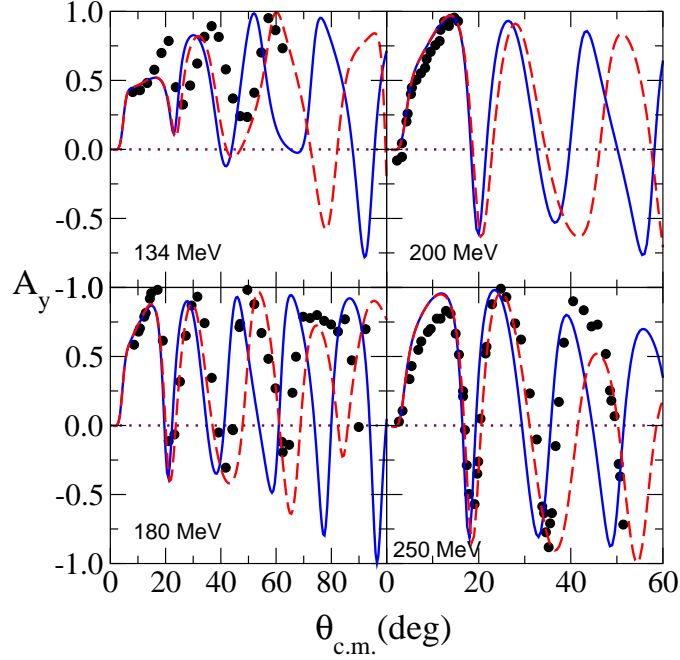}}
\caption{\label{Fig14}(Color online) $g$-folding model
predictions of analyzing powers for the elastic scattering of 134,
180, 200 and 250 MeV protons from ${}^{28}$Si compared with data
~\cite{Ol84, Hi88}.}
\end{figure}
are slightly smaller than what is observed.   There are
no such problems with the comparisons of calculated cross sections   and
data for the higher energies, though, and as found with   other targets,
data and predictions do not match well at large scattering angles.    As
with other targets too,  such discrepancies occur when the cross-section
values are smaller than a few tenths of a mb/sr at most.

In Fig.~\ref{Fig14},  we  present  the  results of the $g$-folding model 
calculations of analyzing powers from proton elastic scattering    from
${}^{28}$Si   at   four   incident   energies   at   which data has been 
taken~\cite{Ol84,Hi88}. The notation used is that as in Fig.~\ref{Fig13}.
The results  for  134  MeV  proton scattering do not match the data well 
though the general trend of the data is seen in the results,  whether WS
or SkX model structures are considered. At the higher energies,  the SHF
model results do match data quite well at least to scattering angles for
which the cross sections are $> 1$ mb/sr;  though there is a mismatch at
forward scattering angles with the 250 MeV data.  The main effect of the
choice of single-nucleon bound-state wave functions, as seen previously,
is again evident with the structure in the WS results spread to   larger
values of momentum transfer than  found using  the SkX model functions. 

\begin{figure}
\scalebox{0.50}{\includegraphics*{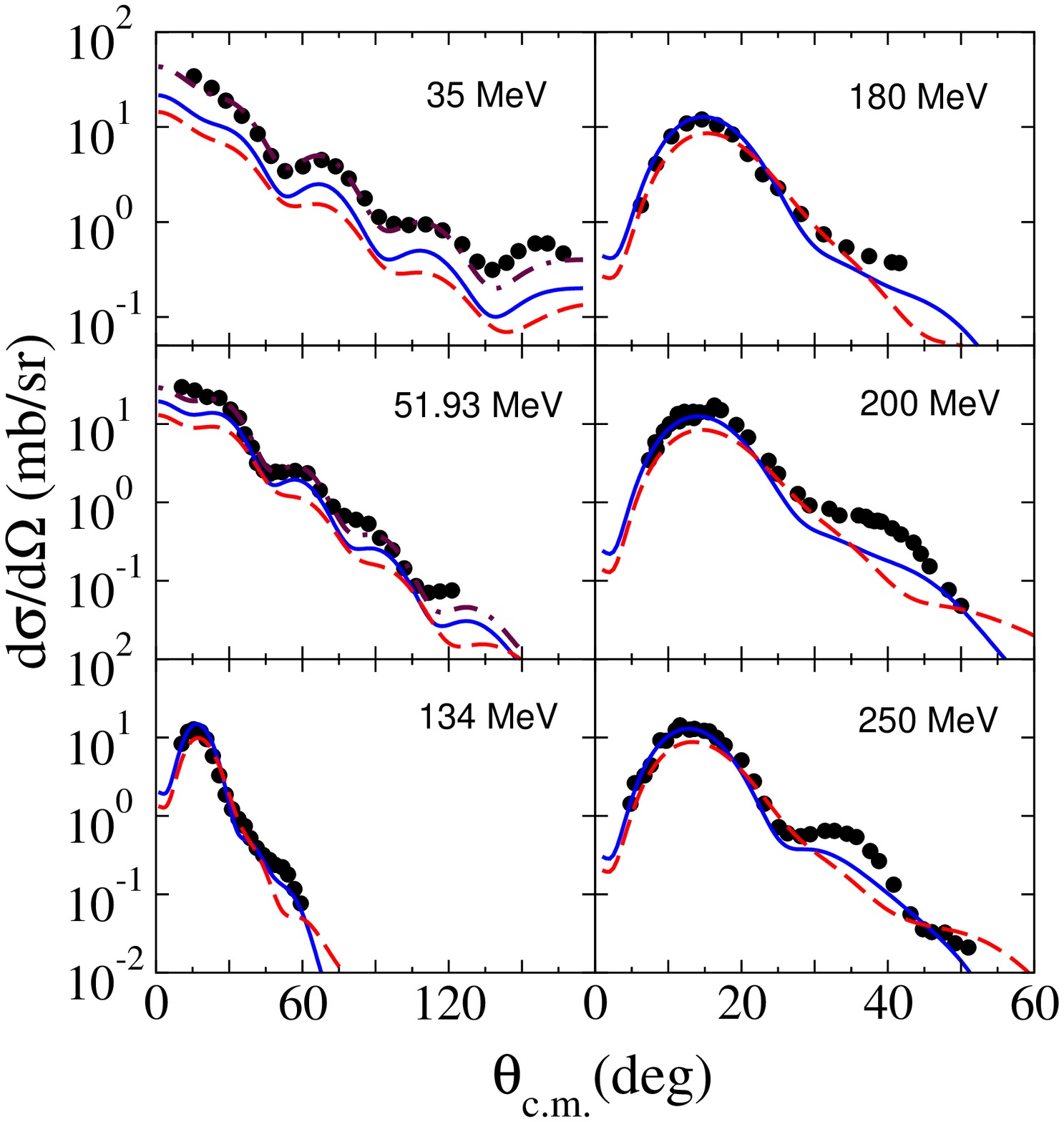}}
\caption{\label{Fig15}(Color online) DWA cross sections for
the inelastic scattering of 35, 51.93, 134, 180, 200 and 250 MeV
protons from the excitation of the $2^+$;1.78 MeV state in
${}^{28}$Si compared with data~\cite{Pi86,Oh80,Hi88,Ch90}.
}
\end{figure}
\begin{figure}
\scalebox{0.50}{\includegraphics*{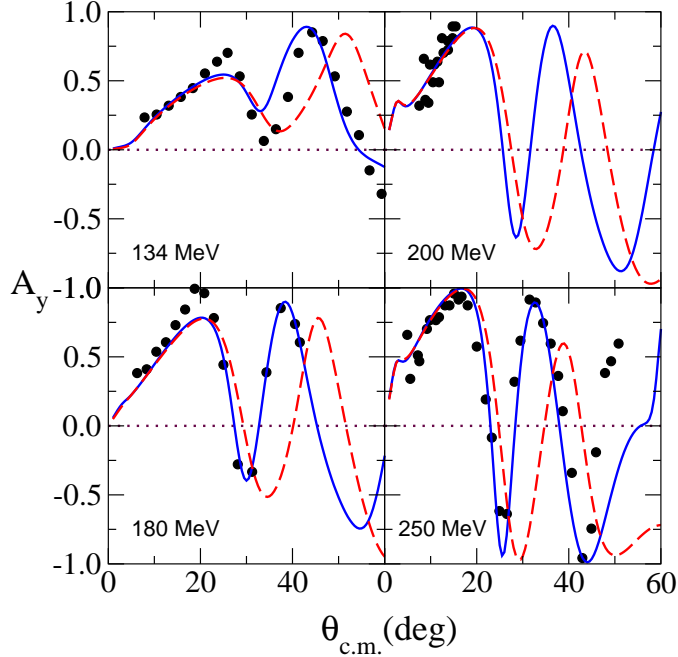}}
\caption{\label{Fig16}(Color online) Analyzing powers for the
inelastic scattering of 134, 180, 200 and 250 MeV protons from the
excitation of the $2^+$;1.78 MeV state in ${}^{28}$Si compared with
data~\cite{Hi88,Ch90}.}
\end{figure}

The  cross  sections resulting from DWA calculations of proton inelastic
scattering to the $2^+$(1.78 MeV) state in ${}^{28}$Si are compared with
data~\cite{Pi86, Oh80, Hi88, Ch90} in Fig.~\ref{Fig15}.   Cross sections
for six incident energies are shown  with  the  solid  and dashed curves 
again portraying results obtained by using, respectively, the SHF and WS
sets  of  single-nucleon bound-state wave functions in the calculations.
Both theoretical predictions lie below the data for the 35 and 51.93 MeV
cases. However, quite good agreement with the measured values is   found
when   the   SHF  results  are  multiplied  by  factors  of 2.0 and 1.5, 
respectively.       Those enhanced results are depicted by the dot-dashed 
curves in the figure.    At the higher energies, both model calculations
match well the forward peak in the data and for which the data magnitude
exceeds $\sim 1$ mb/sr.

The analyzing power from inelastic proton scattering  on ${}^{28}$Si and
leading to the  $2^+$(1.78 MeV) state has been measured at four energies.
That data~\cite{Hi88, Ch90} are compared with  the results of our    DWA
calculations in Fig.~\ref{Fig16}.        The notation is that as used in 
Fig.~\ref{Fig15}. The energies are 134, 180, 200, and 250 MeV, for which
the predicted cross sections      (shown in Fig.~\ref{Fig15}) match data 
quite well.    Concomitantly, the SkX model results in particular match 
the measured analyzing powers very well.        However, the distinctive
feature in the  analyzing  power caused by the choice of single-particle 
bound-state wave functions, is again evident.  Clearly the results found
using the SkX set match the analyzing power data but those found   using
the WS set do not.

\subsection{Elastic scattering of protons from ${}^{40}$Ca}

\begin{figure}
\scalebox{0.50}{\includegraphics*{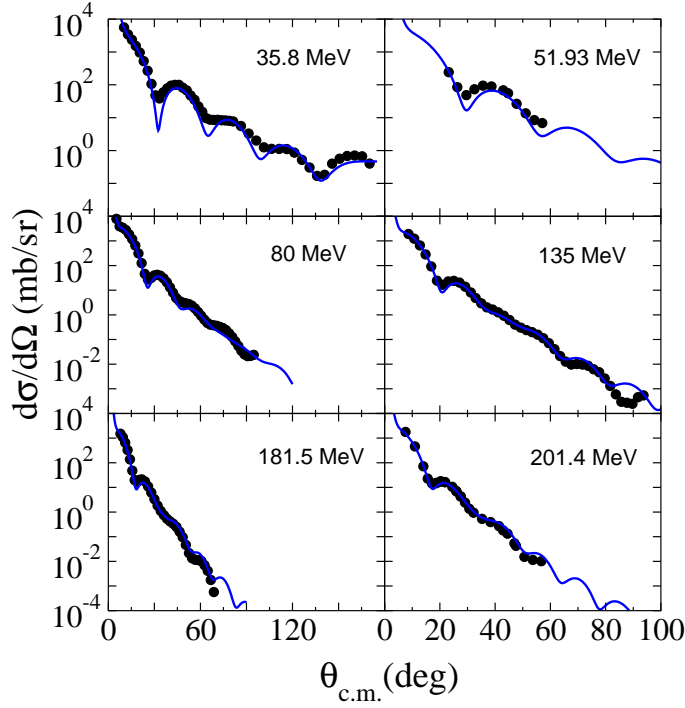}}
\caption{\label{Fig17}(Color online) Differential elastic cross
section predictions of 35.8, 51.93, 80,135, 181.5 and 201.4 MeV
protons scattering from ${}^{40}$Ca compared
with data ~\cite{Gr67,Oh80,Na81,Se93}.}
\end{figure}
\begin{figure}
\scalebox{0.50}{\includegraphics*{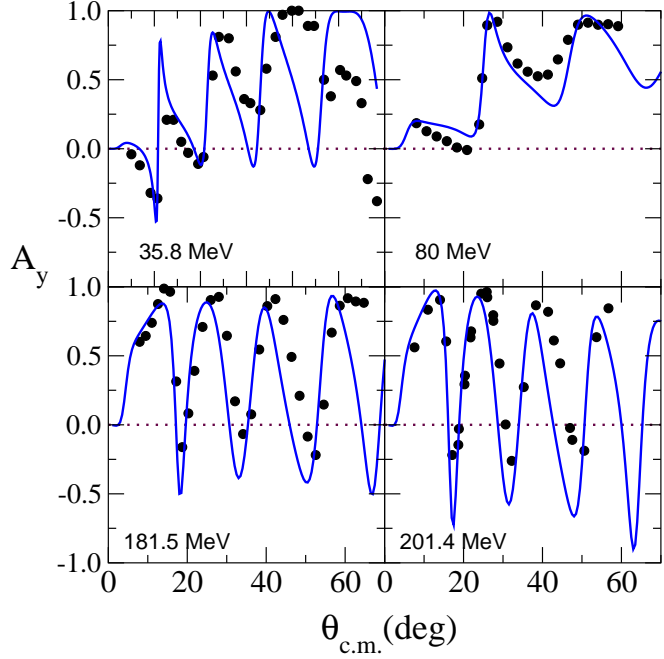}}
\caption{\label{Fig18}(Color online) Analyzing powers
predictions of 35.8, 80, 181.5 and 201.4 MeV protons elastically 
scattered from ${}^{40}$Ca compared with data~\cite{Gr67,Sc82,Se93}.}
\end{figure}

The  results  of  our  calculations, made using the SHF structure in the 
$g$-folding approach, of the elastic scattering of 35.8, 51.93, 80, 135,
181.5, and 201.4 MeV protons from ${}^{40}$Ca are displayed by the solid
curves in Fig.~\ref{Fig17} and compared  with differential cross-section
data~\cite{Gr67,Oh80,Na81,Se93}.    At all energies, there is quite good 
agreement between the predictions and data though there are        more
pronounced minima seen theoretically in the cross sections for 35.8  and
51.93 MeV protons.       For the energies between 80 and 201.4 MeV, our 
predictions  track the data very well even to a magnitude of 0.01 mb/sr.
The quality of  these predictions are reflected in the match between the
associated predictions and the data~\cite{Gr67,Sc82,Se93}        for the
analyzing powers.     As is evident in Fig.~\ref{Fig18}, our $g$-folding
model  results reproduce the observations on the analyzing powers  quite
well.      Notably, the structure falls at the correct momentum transfer 
values, the relative magnitudes of the oscillations are found correctly,
the angle incline trend of the data at   35.8 and  80 MeV is reproduced, 
and the unusual shape feature of the data, once thought to be a 
relativistic effect, is found.

\section{Conclusions}
\label{Conclusions}

Proton scattering data for non-relativistic energies (between 35 and 250
MeV) from  ${}^{12}$C,  ${}^{20}$Ne,  ${}^{24}$Mg,  ${}^{28}$Si      and 
${}^{40}$Ca, have been analyzed to probe the character of the  structure
assumed for those targets.     Differential cross sections and analyzing
power data from both elastic scattering, and, for all bar   ${}^{40}$Ca,
inelastic scattering to the $2^+_1$ states, have been studied.       The
elastic scattering data analyses have been made using a $g$-folding model
for the optical  potentials  in  which  a  complex,  medium  and  energy 
dependent, effective $NN$ interaction has been folded with the full 
one-body density matrices of each target. Three and higher-body interactions
are assumed not to contribute significantly  (in so far as magnitudes of
cross sections are concerned at least).       However, we maintained the
Pauli principle in this approach, and the nonlocal terms in the  optical
potentials  arising  from  ensuring  that  the  Pauli  principle  is not 
violated, have very significant effects.

From those optical potential calculations, the proton-nucleus   relative
motion  wave  functions  also  have been found and used as the distorted 
waves within DWA evaluations of inelastic scattering observables.    The
effective $NN$ interaction used in the $g$-folding to form the  optical
potentials then has been used as the transition operator effecting   the
inelastic scattering events.   In the DWA approach we have used, the one
body (transition) density matrices  for  the  inelastic  scattering  are
required as weightings on the single-particle states involved.     Those
single-particle  states  we  take to be the same ones that we use in the 
$g$-folding.

With the methods and effective $NN$  interactions  proven  credible from 
many past uses, when a structure has been chosen (OBDME,  single-nucleon
bound-state wave functions),      then just one run of the relevant code 
(DWBA98) gives predictions of observables. Comparison of those with data
then allows a critique of the assumed structure.

We  have  chosen a set of nuclei for which  shell and Skyrme-Hartree-Fock 
models of structure have been used with some  success  to  define  their 
ground-state structures.      With ${}^{12}$C in particular, the no-core
shell model  with  a  complete  $(0+2)\hbar\omega$ basis not only gave a 
good description of the positive parity spectrum,        but also of the 
electromagnetic properties and electron scattering form factors.     The
Skyrme-Hartree-Fock   studies  when  constrained  against    shell-model 
occupancies gave very good electron   form   factors for the $sd$-shell 
nuclei that we consider.          While the shell model also provided the 
structure details for excitation of the $2^+_1$ state in ${}^{12}$C,  we
resorted to using projected Hartree-Fock models to specify the OBDME  for
the excitations of the $2^+_1$ states in ${}^{20}$Ne, ${}^{24}$Mg,   and
${}^{28}$Si,   since no other model we know of yet can give estimates of 
those items in the size of basis we wished to consider.              The 
single-particle  bound-state  wave  functions  were  chosen to be either 
oscillator or Woods-Saxon in form for the evaluations with ${}^{12}$C as
target, and of either Woods-Saxon form or the canonical set used in each
SHF evaluation with the other targets.

With the complete $(0+2)\hbar\omega$ space no-core shell model structure
for ${}^{12}$C, our $g$-folding model evaluations of elastic  scattering
cross sections and analyzing powers  for protons with incident  energies 
35, 51.93, 120, 160, 200, and 250 MeV all match data very well.      The
distinctions between using oscillator or Woods-Saxon bound-state    wave
functions were slight.  With the lowest two energies,    there were more
severe minima in the theoretical predictions than in the relevant  data,
but overall the results matched data well,    especially  for scattering
angles where the data magnitude was greater than a few tenths of a mb/sr.
The inelastic  scattering results did not give as good a match to   that
cross-section data, but,     with the exception of the results at 35 and 
51.93 MeV, the predictions matched the dominant forward angle peaks   in
shape and magnitude. Only when data were small in magnitude, as they are
at larger scattering angles,   are discrepancies with predictions noted.
Such is emphasized in the comparisons between predictions and data   for
the analyzing powers (data taken only at 120, 160, 200, and 250 MeV). 

It is important  that  the correct transition strength is found from our
calculations.   This correlates with the close matching of the evaluated 
$B(E2)$ value from the chosen spectroscopy to the observed  $\gamma$-decay
and to  the very good results found on  using  the  same  structure   in
evaluating electron scattering form factors. The under-prediction of the
inelastic scattering cross sections at 35 MeV and,  to  a lesser extent,
at 51.93 MeV then  signals  an  effect  of  a  process additional to the 
reaction mechanism we have assumed.    The back angle bump in the 35 MeV
inelastic scattering  cross-section  data  indicates  that  the  missing
elements are due to virtual excitation of isoscalar $E2$ and/or $E3$   
giant resonances. 

There  is  little  data  available  for  ${}^{20}$Ne,  and with that our 
results give some ambivalence for interpretation.    At 35.2 MeV only an 
average match is made to (elastic scattering)  cross-section  data  that
exceed 1 mb/sr.  Perhaps there is an indication of virtual excitation of 
a giant resonance,  but data of more precision and at other energies are
needed to confirm such.   The 135.4 MeV (elastic) cross-section data are
very well described by the result from the $g$-folding model in which the
SHF wave functions were used.    But the analyzing power found from that
calculation does not match the observed values.     Inelastic scattering
data, both cross section and analyzing power,    have been measured with 
${}^{20}$Ne but only at 135.4 MeV.    With these data however, using the
DWA with OBDME from a PHF model     (with SkX model single-particle wave 
functions) gave   very good results   when  compared  with  the   shapes
of both cross-section and analyzing power data.  Using WS wave functions
did not give as good values.  However, the bare prediction for the cross
section had to be enhanced by 40\% to meet the data.   Interpreting that
as evidence for more correlations  (as a core polarization)  equates  to 
needing a polarization charge of just $0.1e$.

The elastic scattering cross sections  from  ${}^{24}$Mg  also  are very 
well reproduced by predictions made using the $g$-folding model and the SHF
set of wave functions.  The results are preferable to  those found using
the WS functions but the discriminations are only by comparison of  high
momentum, small value, parts of the cross sections. The 34.9 MeV results
have sharper structure that in the data.          The elastic scattering
analyzing powers, taken at 134.7 and 250 MeV,       match the data  only
reasonably,   and that at scattering angles for which the cross sections
are larger than a mb/sr.

Using  the  PHF set of OBDME with the SHF single-particle wave functions 
in DWA calculations,        underestimate the cross sections from proton 
inelastic scattering to the $2^+_1$ state in ${}^{24}$Mg.  That is so at
all four energies considered.           The 51.93 MeV result is slightly
exceptional to those at the other energies    requiring an enhancement of 
2.5 rather than the (average one of) 1.6.          Treating this average 
enhancement as a core polarization   requirement  to  the  PHF structure 
assumed, equates to an effective charge of $0.13e$. Analyzing power data
exist only for  250  MeV  and  the  DWA  result found using the SHF wave 
functions match  that data well to $40^\circ$ by which scattering angle, 
the cross section is but a tenth of a mb/sr.

Much  the  same  as with ${}^{24}$Mg is noted in regard to comparison of 
our $g$-folding model predictions of the elastic  scattering observables
from ${}^{28}$Si.      The sharper variation of predicted cross sections
compared with the 35 and 51.93 MeV data is most obvious.    However, the
(elastic) analyzing power data from ${}^{28}$Si are better replicated by
our predictions than was the case with ${}^{24}$Mg;   though the results
at 135 MeV have larger discrepancy than is pleasing. Using the PHF model
OBDME for the excitation of the    $2^+_1$ state in ${}^{28}$Si with the
SHF  wave  functions  in  DWA  evaluations  gave very good cross-section
shapes in comparison with data taken at energies of 35, 51.93, 134, 180,
200, and 250 MeV.   At least that was so for data that exceeded $\sim 1$
mb/sr.   The match to the magnitudes also was good save that the results
at  35  and  51.93  MeV  required  enhancing   (by a factor of 2 and 1.5 
respectively). Given that no such enhancement was required to match data
at the higher energies, and given that the effective interactions at all
energies considered seem well established,  it is tempting to view these 
35 and 51.93 MeV results as suggestive of virtual excitation of    giant
resonance effects.      Much more data are needed to make such more than 
speculation however.

Finally we considered the data  from  proton  elastic  scattering   from 
${}^{40}$Ca. Over the range of energies studied,   the data,  especially 
the analyzing powers,  show considerable variation.    That variation is
remarkably well defined by our predictions made using the SHF  structure 
model giving much credibility to the effective $NN$ force we have used.

These results serve as indicators to what is needed when analyzing  data 
from RIB scattering from hydrogen (and in fact from any other light mass 
target)   are   taken at energies within the range   of giant  resonance  
excitation  of  the  exotic ion, if such should exist.  Furthermore, our 
results have shown how important it is,       not only to use a credible 
theory of scattering but also one of structure.   Reaction processes not 
considered, effects due to inherent violation of  the  Pauli principle in 
scattering, and inadequate structure, possibly even equating  to  just a
need for the ubiquitous $0.5e$ polarization charge,     may be masked by 
judicious choice of parameter values and  arbitrary scale factors.   The 
more such have to be used, the less physics the analysis can yield.

\begin{acknowledgments}
This research was supported by a 2007 research grant of the Cheju 
National University and by the National Research Foundation of South 
Africa.
\end{acknowledgments}


\bibliography{Kim-C-Ca}

\end{document}